\documentclass{PoS}

\def\ms{M$_{\odot}$}
\def\zs{Z$_{\odot}$}

\title{Topics on Galactic Chemical Evolution}

\ShortTitle{Galactic Chemical Evolution}

\author{\speaker{Nikos Prantzos}\\
        UMR7095 UPMC and Institut d'Astrophysique de Paris\\
        E-mail: \email{prantzos@iap.fr}}

\abstract{I discuss three different topics in Galactic chemical evolution:
          the "puzzling" absence of any observational signature of
          secondary elements ; the building of the Galactic halo in the
          framework of hierarchical galaxy formation, as evidenced from its
          metallicity distribution ; and the potentially important role
          that radial migration may play in the evolution of galactic disks,
          according to recent studies. 
          }

\FullConference{11th Symposium on Nuclei in the Cosmos, NIC XI\\
		July 19-23, 2010\\
		Heidelberg, Germany}

\begin{document}

\section{The elusive secondary elements}

In the field of galactic chemical evolution {\it primary} elements are those produced
from the initial (essentially primordial) H and He entering a star at its formation, while
{\it secondary elements} are those formed from all other elements (metals) entering
the star. Thus, the  {\it yields} of primary elements are independent of the stellar
metallicity, while those of secondaries increase with it. As a result, the secondary/primary
ratio is expected to increase linearly with metallicity. Typical examples of primaries
are the $\alpha$ nuclides ($^{12}$C, $^{16}$O, $^{20}$Ne, etc.) as well as $^{56}$Fe,
while it was traditionally thought that $^{14}$N (produced from initial C and O
through the CNO cycle in H-burning), or the s-elements (produced by n-captures
on seed Fe nuclei) are typical secondaries. It turns out, however, that neither those
elements, nor any other (up to now) shows the expected typical behaviour of secondaries.
In other terms, the concept of secondary element remains only theoretical up to now,
with no observational substantiation. 
In some cases, we (think that we) understand the relevant
observations, but in others the situation is still unclear.

\subsection{The quest for primary Nitrogen}

The behaviour of N as primary (i.e. [N/Fe]$\sim$0) was known for sometime, but it was recently confirmed from VLT mesurements (Spite et al. 2005) down to the realm of very low metallicities ([Fe/H]$\sim$-3, see Fig. 1 middle right). For a long time, the only known source of primary N was Hot Bottom Burning (HBB) in massive AGB stars. Such stars
 (typical mass $M\sim$8 \ms) have lifetimes $\sim$10$^8$ yr, considerably longer than those of typical SNII progenitors (20 \ms \ stars living for $\sim$10$^7$ yr); thus, it is improbable, albeit not impossible\footnote{The timescales of the early Galactic evolution are not constrained (there is no age-metallicity relation) and the contribution of AGBs to chemical enrichment 
even as early as [Fe/H]$\sim$--3 cannot be absolutely excluded.} that that they contributed to the earliest enrichment of the Galaxy with N. On the other hand, massive stars were thought to produce N only as secondary (from the {\it initial} CNO) and not to be at the origin of the observed behaviour.

Rotationally induced mixing in massive stars changed the situation considerably: N is now produced by H-burning of C and O {\it produced inside the star}. As in the case of HBB in massive AGBs,
 N is produced after mixing of protons in He-rich zones, where $^{12}$C originates from the 3-$\alpha$ reaction, i.e. N is produced as primary; it is subsequently ejected to the ISM mostly by the winds of those massive stars.   Stellar models rotating at 300 km/s (typical velocity for solar metallicity stars)
 at all metallicities, did not provide enough primary N at low metallicities to explain the data (Prantzos 2003a and Fig. 1, middle right). Assuming that low metallicity massive stars were rotating faster than their high-metallicity present-day counterparts  (at 800 km/s) leads to a large production of primary N, even at low $Z$  and allows one to explain the data (e.g. Chiappini et al. 2006 and Fig. 1, middle right). Thus, there appears to be a "natural" solution to the problem of early primary N, which may impact on other isotopes as well (e.g. $^{13}$C, produced in a similar way). Even more important, it may also impact   on the next item in the list, namely the evolution of beryllium.

\begin{figure}
\begin{center}
\includegraphics[width=0.4\textwidth]{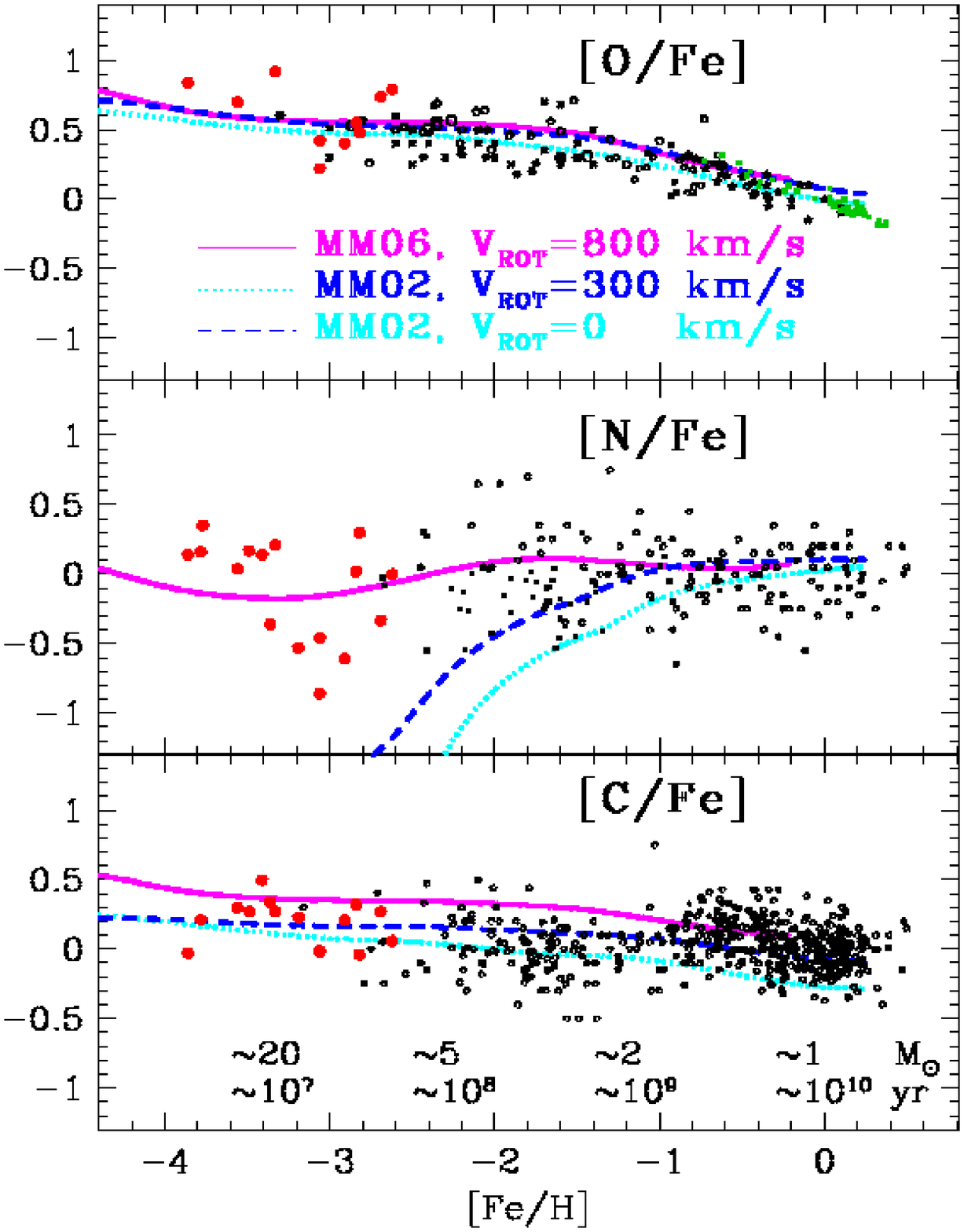}
\qquad
\includegraphics[width=0.4\textwidth]{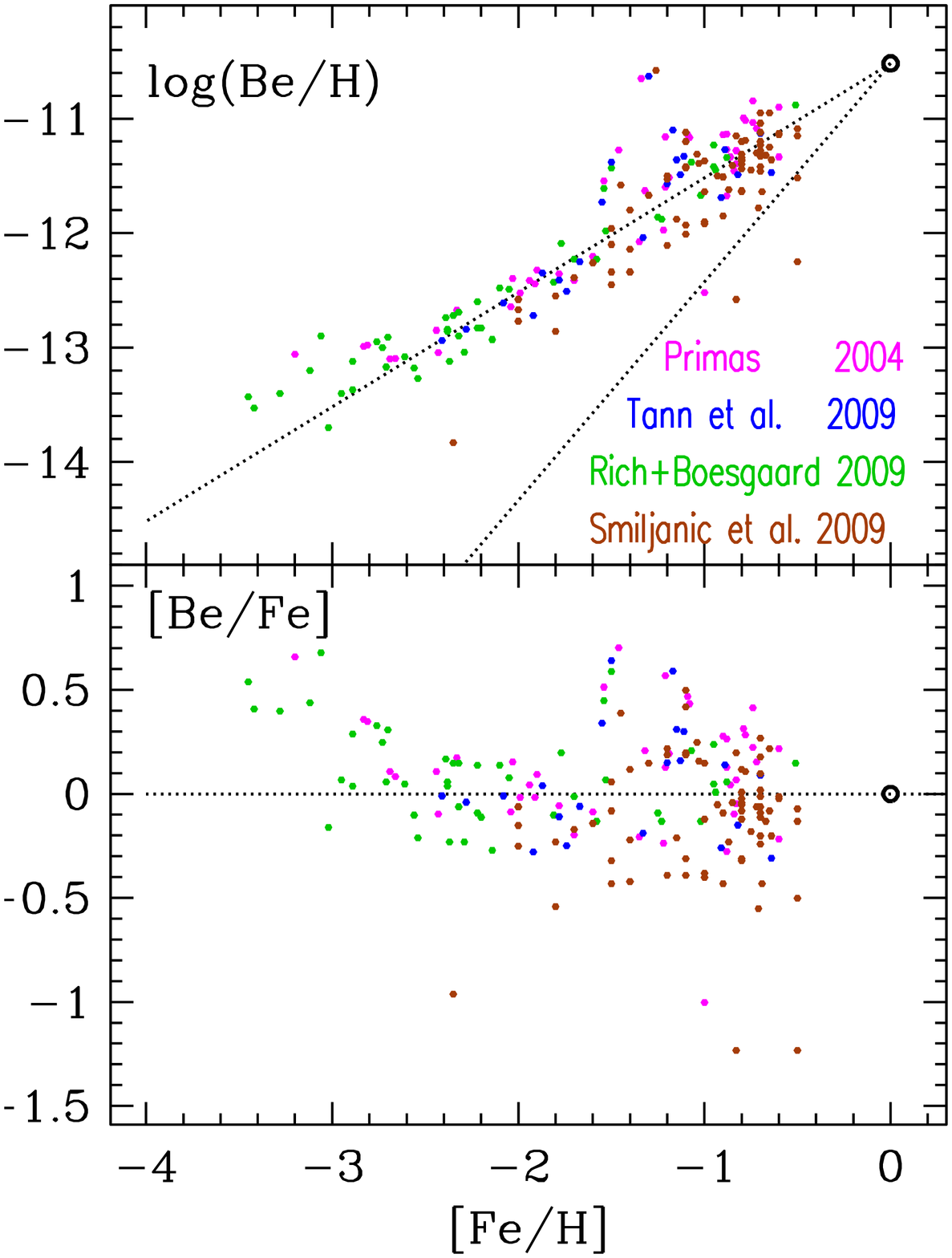}
\caption{{\bf Left:} Evolution of C, N and O vs Fe/H. Solid (purple) curves 
correspond to yields of fast rotating stars at low metallicity and reproduce the
observed early behaviour of N. Evolutionary timescales corresponding to the metallicity
scale and masses of stars dying in those timescales appear in the lower panel.{\bf
 Right}: Observations of Be/H and Be/Fe vs Fe/H; the two dotted lines
 correspond to primary and secondary behaviour, respectively.   }
\label{fig:0}
\end{center}
\end{figure}

\subsection{The quest for primary Be}

Observations of halo stars in the 90s revealed a linear relationship
between Be/H and Fe/H (Gilmore et al. 1991, Ryan et al. 1992) 
as well as between B/H and Fe/H (Duncan et al. 1992). 
That was unexpected, since Be and B
were thought to be produced as {\it secondaries}, by spallation of the 
increasingly abundant CNO nuclei of the ISM during the propagation of 
protons and aplhas of Galactic Cosmic rays (GCR). 
The only way to produce primary Be  is
by assuming that it is produced by the fragmentation of  the CNO nuclei of GCR,
as they hit the p and $\alpha$ of the ISM and that GCR 
have always the same CNO content (Duncan et al. 1992);
other efforts to enhance the early production of Be, by e.g. 
invoking a better confinement - 
and thus, higher fluxes - of GCR in the early Galaxy (Prantzos et al. 1993)  only partially succeeded\footnote{The observed primary evolution of B can be explained
by assuming $\nu$-induced production of its major isotope $^{11}$B in core collapse
SN (Olive et al. 1994).}.
 The reason 
was clearly revealed by the ``energetics argument"
put forward by Ramaty et al. (1997): if SN are the main source of GCR energy,
there is a limit to the amount of light elements produced per SN, which depends on GCR and ISM composition. If the metal content of {\it both} ISM and GCR is
low, there is simply not enough energy
in GCR to keep the Be yields constant. The only possibility
to have $\sim$constant LiBeB yields is by assuming that the ``reverse" 
component of GCR (fast CNO nuclei) is primary,
i.e. that GCR have a $\sim$constant metallicity (Fig. 2 in Prantzos 2010). 
This has profound implications for our 
understanding of the GCR origin. It should be noted that before those observations,
no one would have the idea to ask ``what was the GCR composition in the early Galaxy?".

For quite some time it was thought that GCR originate from the average ISM, where they
are accelerated by the {\it forward shocks} of SN explosions; this can only produce secondary Be.  
A $\sim$constant abundance of C and O in GCR  can ``naturally" be understood if SN
accelerate their own ejecta, trough their {\it reverse schock} (Ramaty et al. 1997).  
However, the absence of unstable $^{59}$Ni 
(decaying through e$^-$ capture
within 10$^5$ yr) from observed GCR suggests that acceleration occurs 
$>$10$^5$ yr after the explosion (Wiedenbeck et al. 1999) 
when SN ejecta are presumably  already diluted in the ISM. 
% Furthermore, the reverse shock has only a small fraction of the SN kinetic energy, while observed GCR require a large fraction of it.
%\footnote{The power of GCR is estimated to $\sim$10$^{41}$ erg s$^{-1}$ galaxywide, i.e. about 10\% of the kinetic energy of SN, which is $\sim$10$^{42}$ erg s$^{-1}$ (assuming 3 SN/century for the Milky Way, each one endowed with an average kinetic energy of 1.5 10$^{51}$ ergs).}.

Higdon et al. (1998) suggested  that GCR are accelerated out of  {\it superbubbles} (SB) material, enriched by the ejecta of many SN as to have a large and $\sim$constant metallicity. In this scenario, it is the forward shocks of SN that accelerate material ejected from other, previously exploded SN (see  
Binns et al. 2005, Rauch et al. 2009).
%That scenario has also been  invoked in order to explain the present day source isotopic  composition of GCR (Binns et al. 2005, Rauch et al. 2009). 
%Notice that the main feature of that composition, namely
%a large $^{22}$Ne/$^{20}$Ne ratio, is explained as due to the contribution of winds from Wolf-Rayet (WR) stars
%(e.g. Prantzos et al. 1987), and the SB scenario offers a plausible (but not unique) framework in bringing together
%contributions from both  SN and WR stars.
The SB scenario suffers from several drawbacks (Prantzos 2010)
which, however,  may not be lethal. Still, it is hard to imagine that SB have always the same average metallicity, especially during the
 early Galaxy evolution, where metals were easily expelled out of the shallow potential wells of the
 small sub-units forming the Galactic halo.

\begin{figure}
\centering
% Use the relevant command to insert your figure file.
% For example, with the graphicx package use
\includegraphics[angle=-90,width=0.82\textwidth]{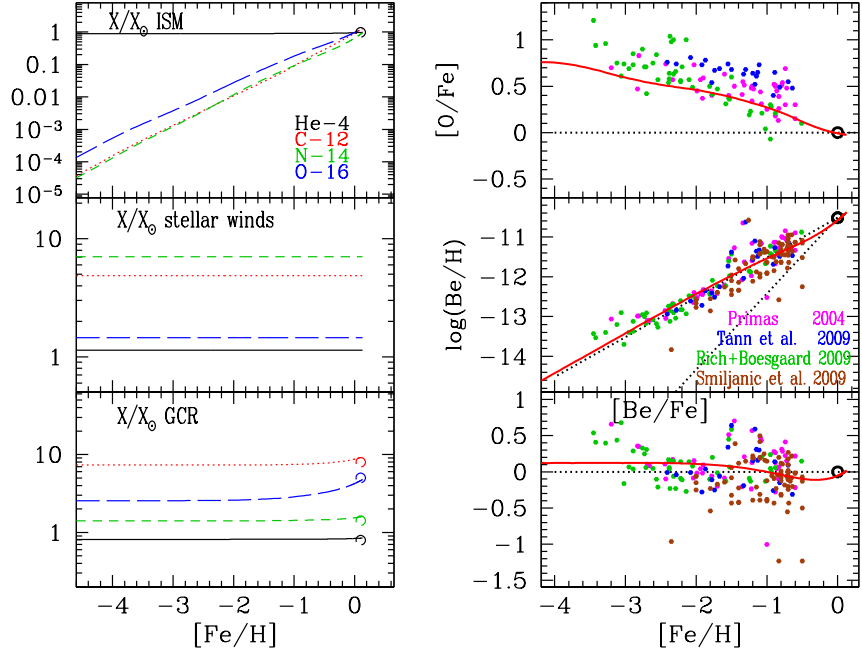}
\caption{{\bf Left:} Evolution of the chemical composition (in corresponding 
solar abundances) of He-4 ({\it solid}), C-12 ({\it dotted}), N ({\it short dashed})
and O ({\it long dashed})in: ISM ({\it top}), massive star winds ({\it middle}) and
GCR ({\it bottom}). {\it Dots} in lower panel indicate estimated GCR source composition
(from Elison et al. 1997). {\bf Right}: Evolution ({\it solid curves} of O/Fe ({\it top}),
Be/H ({\it middle}) and Be/Fe ({\it bottom}); {\it dotted lines} indicate solar values in
top and bottom panels, primary and secondary Be in middle panel.}
\label{fig:2}       % Give a unique label
\end{figure}

A  different explanation for the origin of GCR,   is proposed in Prantzos (2010). 
He notices that {\it rotating} massive stars display 
substantial mass loss down at very low (or even zero) metallicities (see previous section).
%The winds of those stars are enriched in CNO (products of H and He burning {\it within} the star itself).
%at all metallicities and at about the same level; it is precisely this enrichment of the WR winds at
%all metallicities that allows us to understand the observed primary behaviour of N down to the lowest
%metallicity halo stars (Sec. 1.1). This gives some confidence in using the same model results%
%to predict the composition of GCR over the history of the Milky Way.
Assuming  that GCR are accelerated when the forward shocks of SN propagate into the  previously ejected envelopes of rotating massive stars (partially mixed with the  surrounding ISM), one may then 
calculate the evolution of the ISM and GCR composition (Fig. 2, left). It is found that the resulting
Be evolution  nicely fits the data (Fig. 2, right); it is the first time that such a 
calculation is performed {\it not by assuming} a given 
GCR composition, but by {\it calculating} it in
a (hopefully) realistic way.

\begin{figure}
\begin{center}
\includegraphics[width=0.42\textwidth]{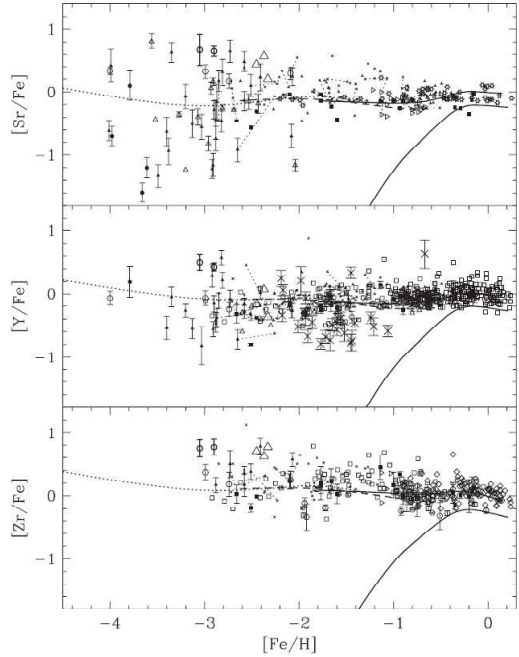}
\caption{Evolution of Sr, Y and Zr in the Galaxy (halo and solar neighborhood): observations
vs a model involving s-process from low and intermediate mass stars ({\it solid} curves)
and a putative n-process ({\it dashed} curves). From Travaglio et al (2004). }
\label{fig:0}
\end{center}
\end{figure}

\subsection{The early appearance of s-elements}

The products of slow n-capture (s-process) were traditionally thought to behave as secondaries. However, various theoretical arguments suggest that this cannot be true in most cases and current
uncertainties prevent from making sound theoretical predictions for the behaviour of those
elements.

1) Solar system s-elements have a primary contribution $f_r=Y_r/(Y_r+Y_s)$ (where $Y_s$ and $Y_r$ are
the corresponding yields) from the r-process; 
thus, for an s-element $X$, a "floor" in the [X/Fe] ratio is expected below
some metallicity $Z$ (Truran 1981). But the evolution of $f_r$ with $Z$ is poorly
determined, because of the unknown evolution of $Y_s$ (while $Y_r$ is expected to evolve
roughly as the oxygen yield $Y_O$, at least at late times).

2) $Y_s(Z)$ depends on: i) the behaviour of the "neutron economy trio" (sources - poisons - 
seed nuclei) with metallicity (Prantzos et al. 1990); for instance,
the behaviour of the n-source $^{13}$C($\alpha$,n) 
is different from the one of $^{22}$Ne($\alpha$,n) and 
ii) the mass range of the s-element sources (stars with $M\sim$1.5-3 \ms,
with lifetimes from a few 10$^8$ to a few 10$^9$ yr for the "main" s-component, but
massive stars for the "weak" s-component); since the yields of individual stars $y(M,Z)$ are
unknown, the behaviour of the global yield $Y_s(Z)$ (averaged over the IMF) is unknown also.

In the case of heavy s-elements, like Ba, the observed behaviour of e.g. the
Ba/Eu ratio (Eu being an almost pure r-element) can be explained as resulting from
a pure r- contribution below [Fe/H]$\sim$-1.5 (where [Ba/Eu]$\sim$const.$\sim$-0.6)
and a stronger (but {\it not} monotonically increasing) contribution from the s-process in intermediate mass stars above that value (see Travaglio et al. 1999). However, the 
situation appears much more difficult in the case of the light s-elements
Sr, Y and Zr, which behave {\it exactly} as Fe (i.e. the [X/Fe] ratio is $\sim$constant
down to the lowest metallicities). Since the r- contribution to the solar system abundances of those
elements is small, Travaglio et al. (2004) suggested the operation of an unknown neutron 
capture process (called {\it n-process}) of primary nature in massive stars at low
metallicities. 

One might think that a "natural" site for that process may be 
$^{22}$Ne($\alpha$,n) in
core He-burning in massive {\it rotating} stars: indeed, as stressed in Sec. 1.1,
{\it primary} $^{14}$N is produced in those stars, and the amount remaining in the
He-burning -zones is turned mostly into $primary ^{22}$Ne early in He-burning.  However,
since both the neutron source $^{22}$Ne and the main neutron poisons $^{25}$Mg
and $^{22}$Ne are primary, the s-process in the Sr-Zr region turns out to 
be secondary (scaling with
the Fe seed abundance), as shown in Pignatari et al. (2008) with a 25 \ms \ model
of rotating star at metallicities [Fe/H]=-3 and -4, respectively.
Thus, the observed primary 
behaviour of Sr, Y and Zr at low metallicity remains unexplained at present.

\section{The MW halo in cosmological context}

The metallicity distribution (number of stars per unit metallicity interval) of a galaxian
system gives valuable information about its history, and in particular, the occurence of gaseous
flows (infall, outflow).
 The regular shape of the metallicity distribution of the Milky Way (MW) halo can readily be explained by the simple model of galactic chemical evolution (GCE) with outflow, as suggested by Hartwick (1976). However, that explanation lies within the framework of the monolithic collapse scenario for the formation of the MW (Eggen, Lynden-Bell and Sandage, 1962).
Several attempts to account for the metallicity distribution of the MW halo  in the
modern framework (hierarchical merging of smaller components, hereafter sub-haloes) were undertaken in recent years through numerical simulations  (Bekki and Chiba 2001;    Salvadori et al. 2007).  Independently of their success or failure in reproducing the observations, such models provide little or no physical insight into the physical processes that shaped the metallicity distribution of the MW halo.  Why is its metallicity distribution so well described by the simple model with outflow (which refers to a single system)? And what determines the peak of the metallicity distribution at [Fe/H]=$\sim$--1.6, which is (successfully) interpreted in the simple model by a single parameter (the outflow rate) ?  Here we present an attempt to built the halo metallicity distribution analytically (Prantzos 2008a) in the framework of the hierarchical merging paradigm.

\subsection{The halo metallicity distribution  and the simple model}

The halo metallicity distribution is nicely described by the simple model of GCE, in which the metallicity $Z$ is given as a function of the gas fraction $\mu$ as $ Z \ = \ p \ ln(1/\mu) \ + Z_0 $,
where $Z_0$ is the initial metallicity of the system  and $p$ is the {\it yield} 
(metallicities and yield are expressed in units of the solar metallicity
\zs).  If the system evolves at a
constant mass (closed box), the yield is called the {\it true yield},
otherwise (i.e. in case of mass loss or gain) it is called the {\it
effective yield}. The {\it differential metallicity
distribution} (DMD)  is:
\begin{equation} 
{{d(n/n_1)}\over{d(logZ)}} \ = \
 {{{\rm ln10}}\over{1-{\rm exp}\left(-{{Z_1-Z_0}\over{p}}\right)}}  \
 {{Z-Z_0}\over{p}} 
\ {\rm exp}\left(-{{Z-Z_0}\over{p}}\right)
\end{equation}
where $Z_1$ is the final metallicity of the system and $n_1$ the total
number of stars (having metallicities $\le Z_1$). This function has a
maximum for $Z-Z_0=p$, allowing one to evaluate easily the effective yield $p$ if the DMD is observed.
In the case of  {\it outflow} at a rate $F \ = \ k \ \Psi$ (where $\Psi$ is the Star Formation Rate or SFR)  
one obtains  $ k \ = \ (1-R) \ (p_{True}/p_{Halo} -1)$, where $R\sim$0.35 is the return mass fraction
of the system, $p_{True}$ and $p_{Halo}$ the observationnaly determined yields in the bulge (fitetd with 
a closed box model)
and the halo, respectively. 
The DMD of the MW halo is nicely fit with a simple outflow model with $k\sim$7-8.

\begin{figure}
%\centering
% Use the relevant command to insert your figure file.
% For example, with the graphicx package use
\includegraphics[width=0.46\textwidth]{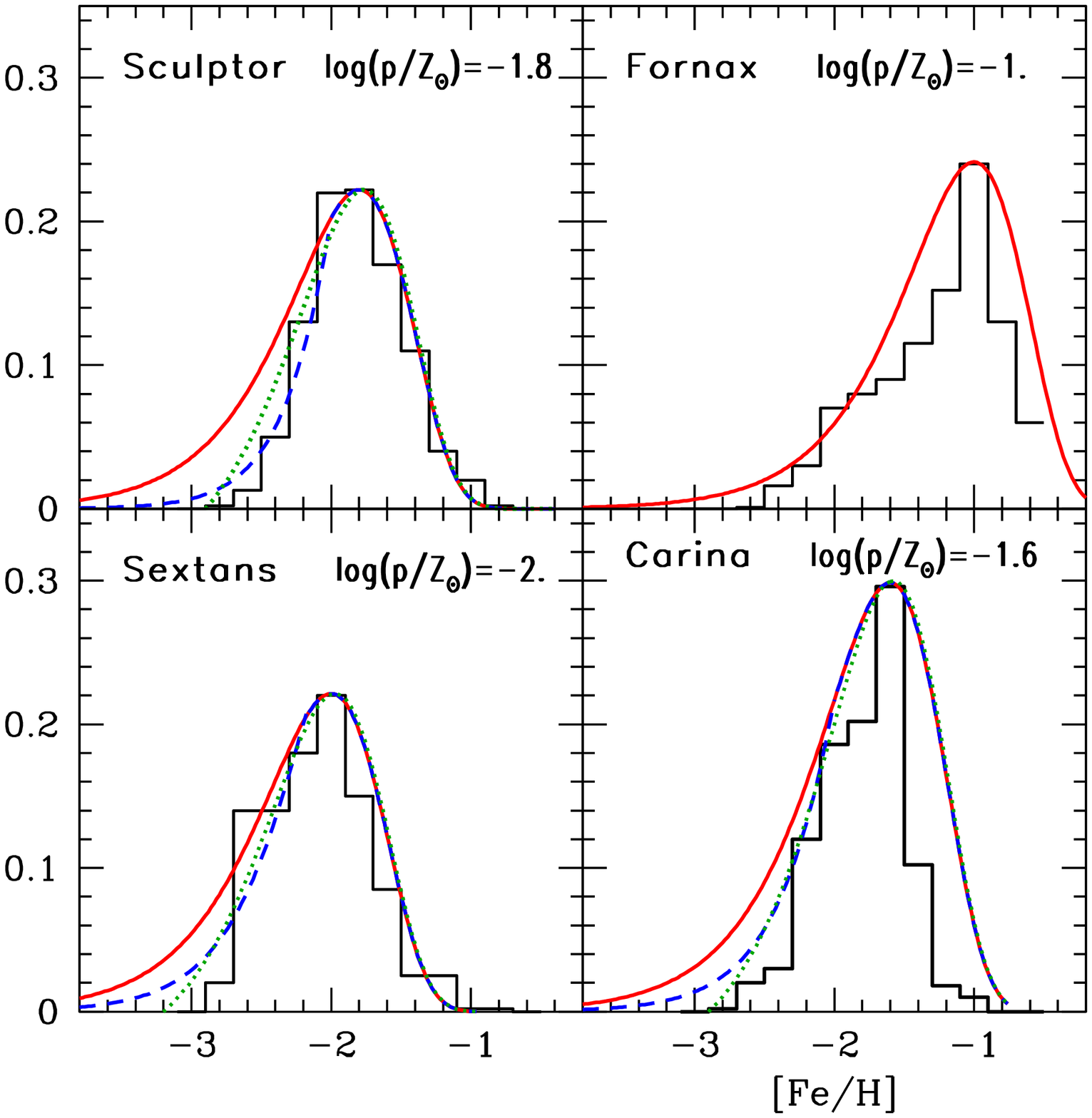}
\qquad
\includegraphics[width=0.46\textwidth]{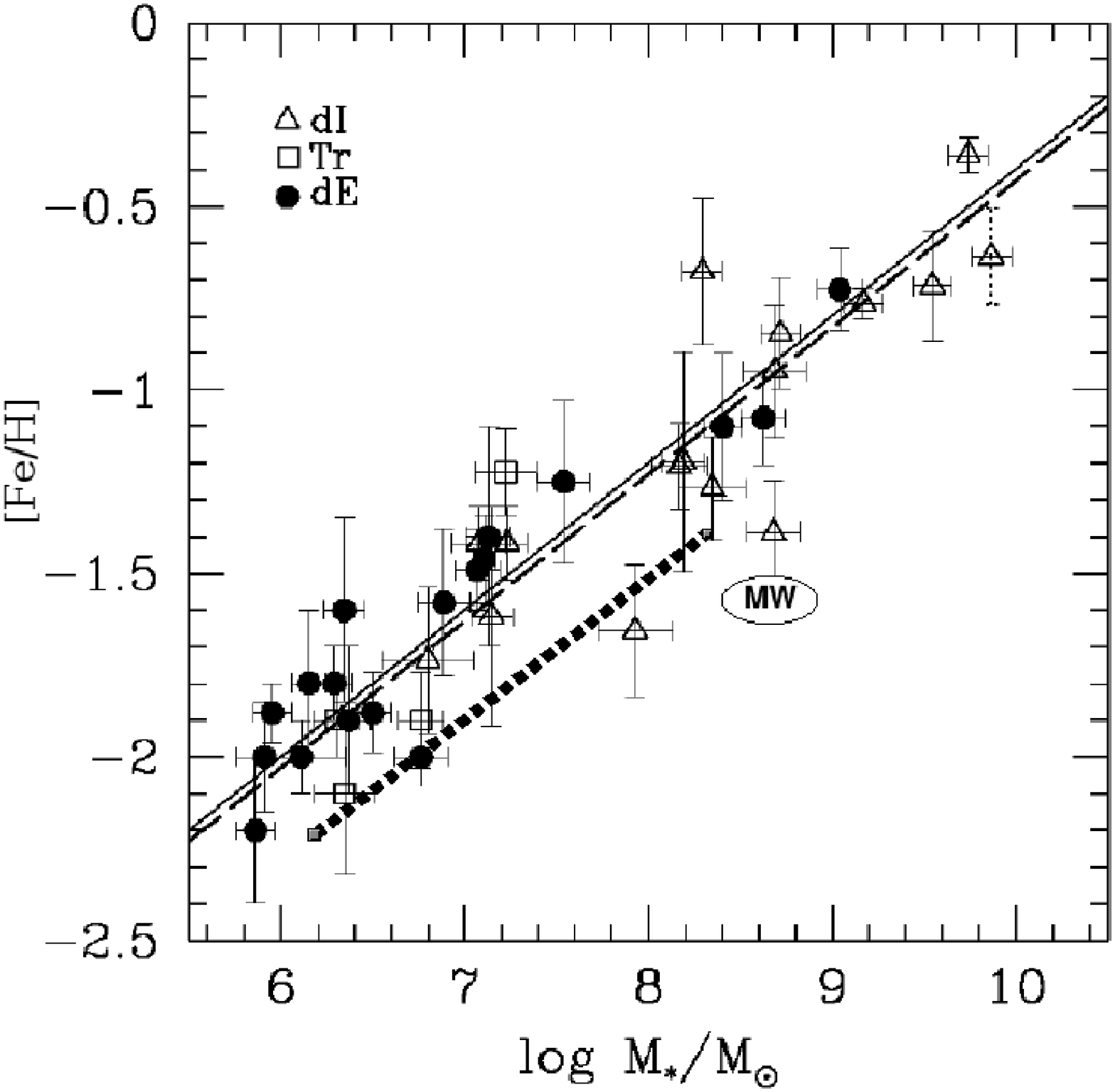}
\caption{{\bf left:} Metallicity distributions  of dwarf satellites of the Milky Way. 
Data are in {\it histograms} (from Helmi et al. 2006). {\it Solid curves} 
indicate the results of simple GCE  models with outflow proportional to 
the star formation rate;  the corresponding effective yields (in \zs) 
appear on top right of each panel. {\it Dashed curves} are fits obtained
 with an early infall phase, while {\it dotted curves} are models with
 an initial metallicity log$(Z_0)\sim$--3; both modifications 
to the simple model
 (i.e. infall and initial metallicity) improve the fits to the data.
{\bf Right:} Stellar metallicity vs stellar mass for nearby galaxies; data
 and model ({\it upper curves}) are from Dekel and Woo (2003), with {\it
 dI} standing for dwarf irregulars and  {\it dE} for dwarf ellipticals. The {\it thick 
dotted} line represents the effective yield of the sub-haloes that formed the MW halo according to this work (i.e. with no contribution from SNIa, see Sec. 3.2). The MW halo, with average metallicity [Fe/H]=--1.6 and 
estimated mass $\sim$4x10$^8$ \ms \ falls below both curves.  }
\label{fig:2}       % Give a unique label
\end{figure}

\begin{figure}
% Use the relevant command to insert your figure file.
% For example, with the graphicx package use
\includegraphics[width=0.46\textwidth]{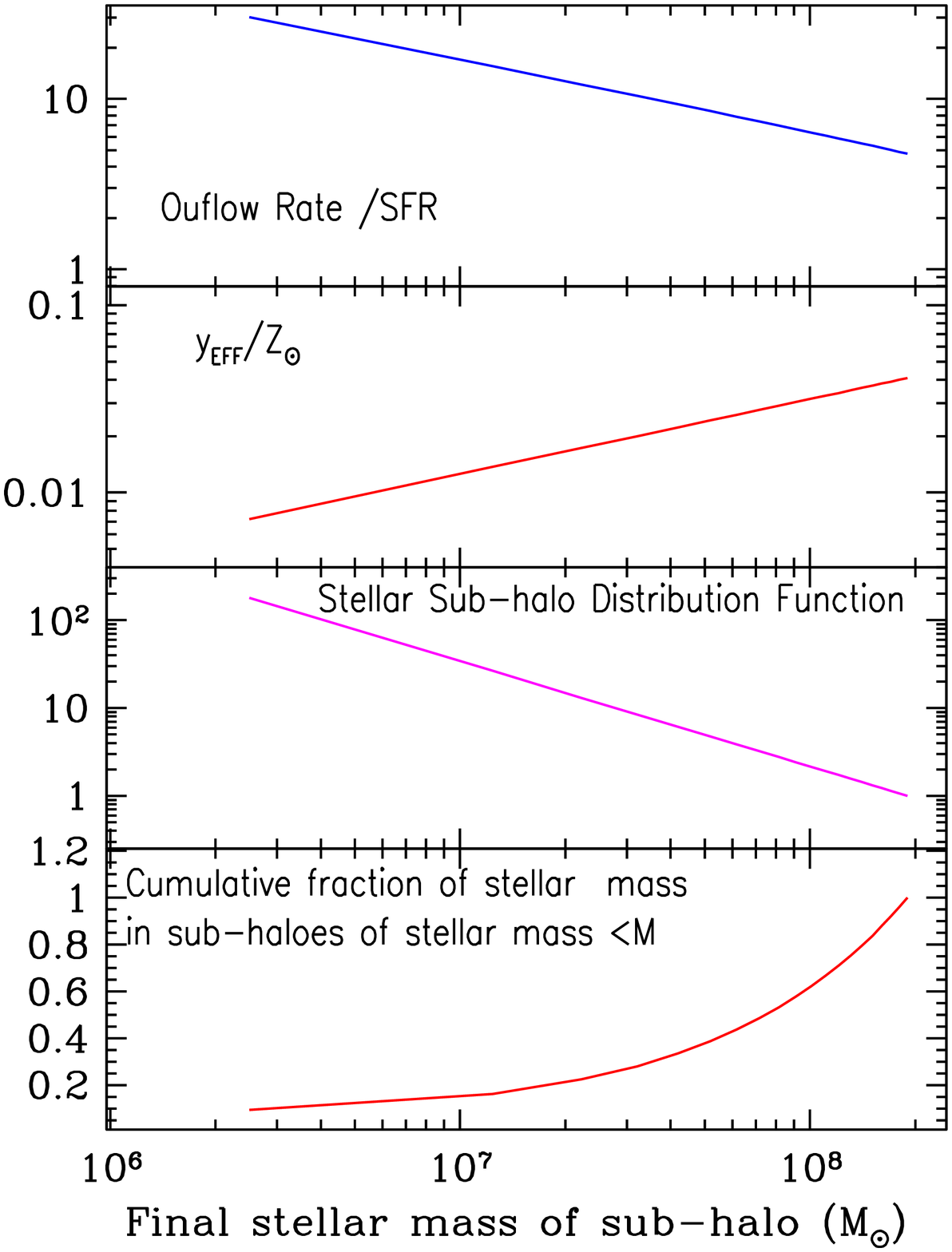}
\qquad
\includegraphics[width=0.46\textwidth]{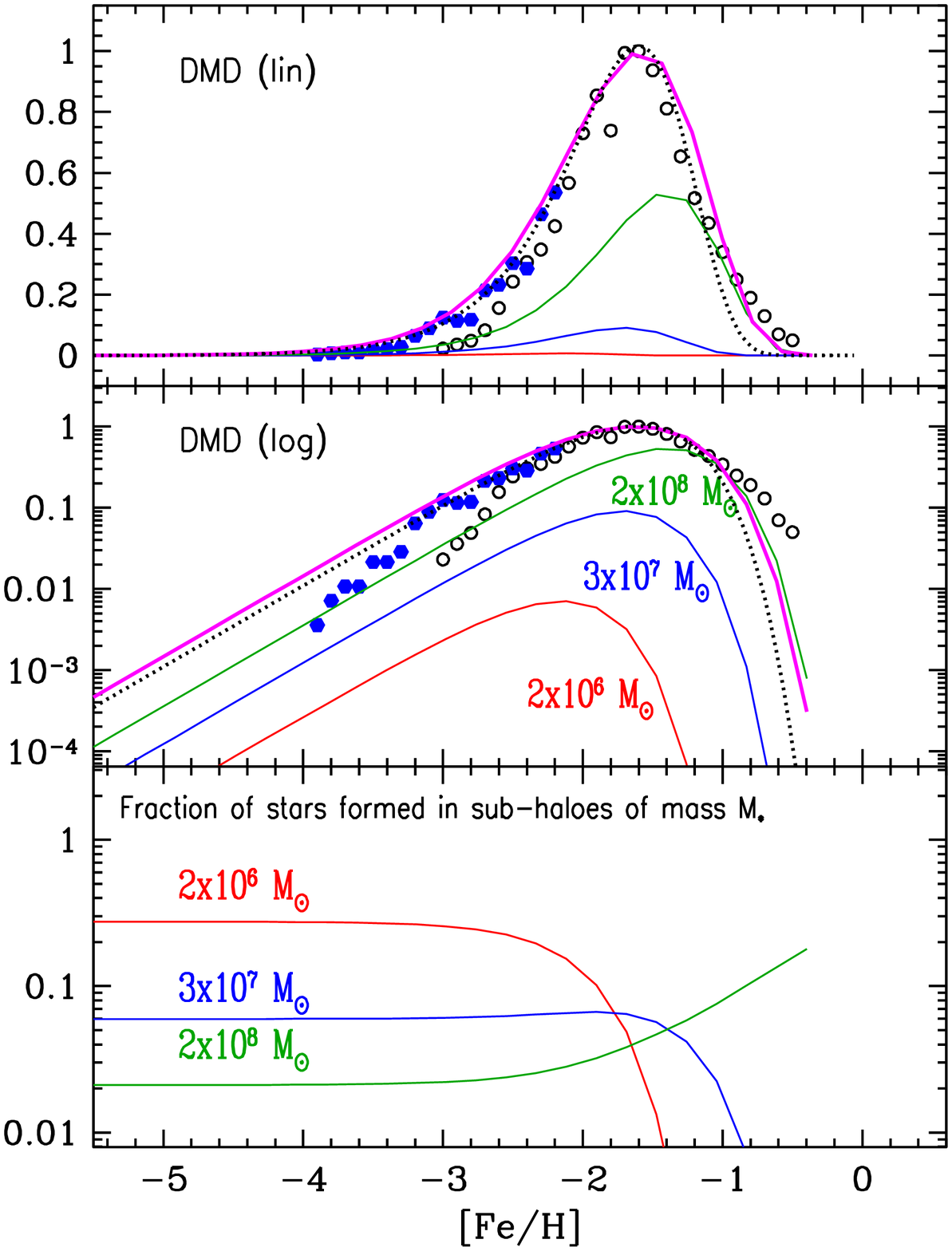}
\caption{{\bf Left:}Properties of the sub-haloes as a function of their stellar
 mass,  empirically derived as discussed
 in Sec. 3. From top to bottom: Outflow rate, in units of the
 corresponding star formation rate; Effective yield, in solar units;
 Distribution function; Cumulative fraction of stellar mass contributed
 by the sub-haloes. The total mass of the MW halo is 4 10$^8$ \ms.  {\bf
 Right}:{\it Top} and {\it middle} panels: Differential metallicity distribution (in lin and log scales, respectively) of the MW halo, assumed to be composed of a population of smaller units (sub-haloes). The individual DMDs of a few sub-haloes, from 10$^6$ \ms \ to 4 10$^7$ \ms, are indicated in the middle panel, as well as the sum over all haloes ({\it solid upper curves} in both panels, compared to observations). {\it Dotted curves} in top and middle panels indicate the results of the simple model with outflow (same as in Fig. 1). Because of their large number, small sub-haloes with low effective yields contribute the largest fraction of the lowest metallicity stars, while large haloes contribute most of the high metallicity stars ({\it bottom panel}).  
 Figure from Prantzos (2008).}
\label{fig:4}
\end{figure}

\subsection{The halo DMD and hierarchical merging}

Assuming  that the MW halo was formed by the merging of
smaller units ("sub-haloes"), one has to know:
a) the DMD of each sub-halo and b) the number distribution of the sub-haloes
 $dN/dM$. Prantzos (2008a) assumed that the DMD of each sub-halo 
 had a DMD described by the simple model with an appropriate effective yield. This
assumption is based on recent observations of the dwarf spheroidal
(dSph) satellites of the Milky Way\footnote{It is
true that the dSphs that {\it we see today} cannot be the components of
the MW halo, because of  their observed abundance patterns (e.g. Shetrone, C\^ot\'e and Sargent 2001;
Venn et al. 2004): their $\alpha$/Fe ratios are typically smaller than
the   [a/Fe]$\sim$0.4$\sim$const. ratio of halo stars. 
This implies that they evolved on longer timescales than the Galactic
halo, allowing SNIa to enrich their ISM with Fe-peak nuclei and thus 
to lower the $\alpha$/Fe ratio by a factor of $\sim$2-3 (as evidenced from the 
[O/Fe]$\sim$0 ratio in their highest metallicity stars).}. 
The DMDs of four nearby dSphs (Helmi et al. 2006) are displayed as histograms in Fig. 4 (left),
where they are compared to the simple model with appropriate effective
yields ({\it solid curves}). The effective yield in each case was simply
assumed to equal the peak metallicity (Eq. 1.1). It can be seen that the overall shape 
of the DMDs is quite well fitted by the simple models. This is important,
since i) it strongly suggests that {\it all} DMDs of small galaxian
systems can be described by the simple model  and ii) it allows to determine {\it
effective yields} by simply taking the peak metallicity of each DMD.
%Before proceeding to the determination of effective yields, we note that the fit of the simple model to the data of dSphs fails in the low metallicity tails. Helmi et al. (2006) attribute this  to a pre-enrichment of the gas out of which the dSphs were formed. However, early infall is another, equally plausible, possibility, as argued in Prantzos (2008). The recent simulations of Salvadori et al. (2008) find both pre-enrichment and early infall for local dSphs.
%If the DMDs of all the components of the Galactic halo are described by
%the simple model, then their shape is essentially  described by the
%corresponding effective yield $p$ (and, to a lesser degree, by the
%corresponding initial metallicity $Z_0$). 
Observations suggest that the
effective yield is a monotonically increasing function of the galaxy's
stellar mass $M_*$ (Fig. 4 right).  In the case of the
progenitor systems of the MW halo, however, the effective yield must have been lower, since SNIa had not time to contribute (as evidenced by the high 
$\alpha$/Fe$\sim$0.4 ratios of halo stars), by a factor of about 2-3. We assume
then that the effective yield of the MW halo components $p(M_*)$ (accreted satellites
or sub-haloes) is given (in \zs) by
%\begin{equation}
%p(M_*) \ = \ 0.005 \ \left({\frac{M_*}{10^6 M_{\odot}}}\right)^{0.4}
%\end{equation}
the thick dotted curve in Fig. 4 (right).
The stellar mass $M_*$ of each of the sub-haloes should be $M_*< M_H$
where $M_H$ is the stellar mass of the MW halo ( $M_H$=
4$\pm$0.8 10$^8$ \ms, e.g. Bell et al. 2007).

Hierarchical galaxy formation scenarios predict  the mass function of
the dark matter sub-haloes which compose a dark matter halo at a given
redshift. Several recent simualtions find
$ dN/dM_D \ \propto \ M_D^{-2}  $ (e.g.  Giocoli et al. 2008).
In our case, we are interested in the mass function of the
{\it stellar sub-haloes}, and not of the dark ones. Considering the effects of outflows on
the baryonic mass function, Prantzos (2008a) finds that $dN/dM_* \ \propto \ M_*^{-1.2}$,
%\begin{equation}
%\frac{dN}{dM_*} \ \propto \ M_*^{-1.2}
%\end{equation}
i.e. {\it the distribution function of the stellar sub-haloes is flatter than
the distribution function of the dark matter sub-haloes}. 
%The normalisation of the stellar sub-halo distribution function is made through
%\begin{equation}
%\int_{M_1}^{M_2} \frac{dN}{dM_*} \ M_* \ dM_* \ = \ M_H
%\end{equation}

\begin{figure}
% Use the relevant command to insert your figure file.
% For example, with the graphicx package use
%\includegraphics[width=0.24\textwidth]{Schorck1.eps} 
%\qquad
%\includegraphics[width=0.24\textwidth]{Schorck2.eps}
\includegraphics[width=0.46\textwidth]{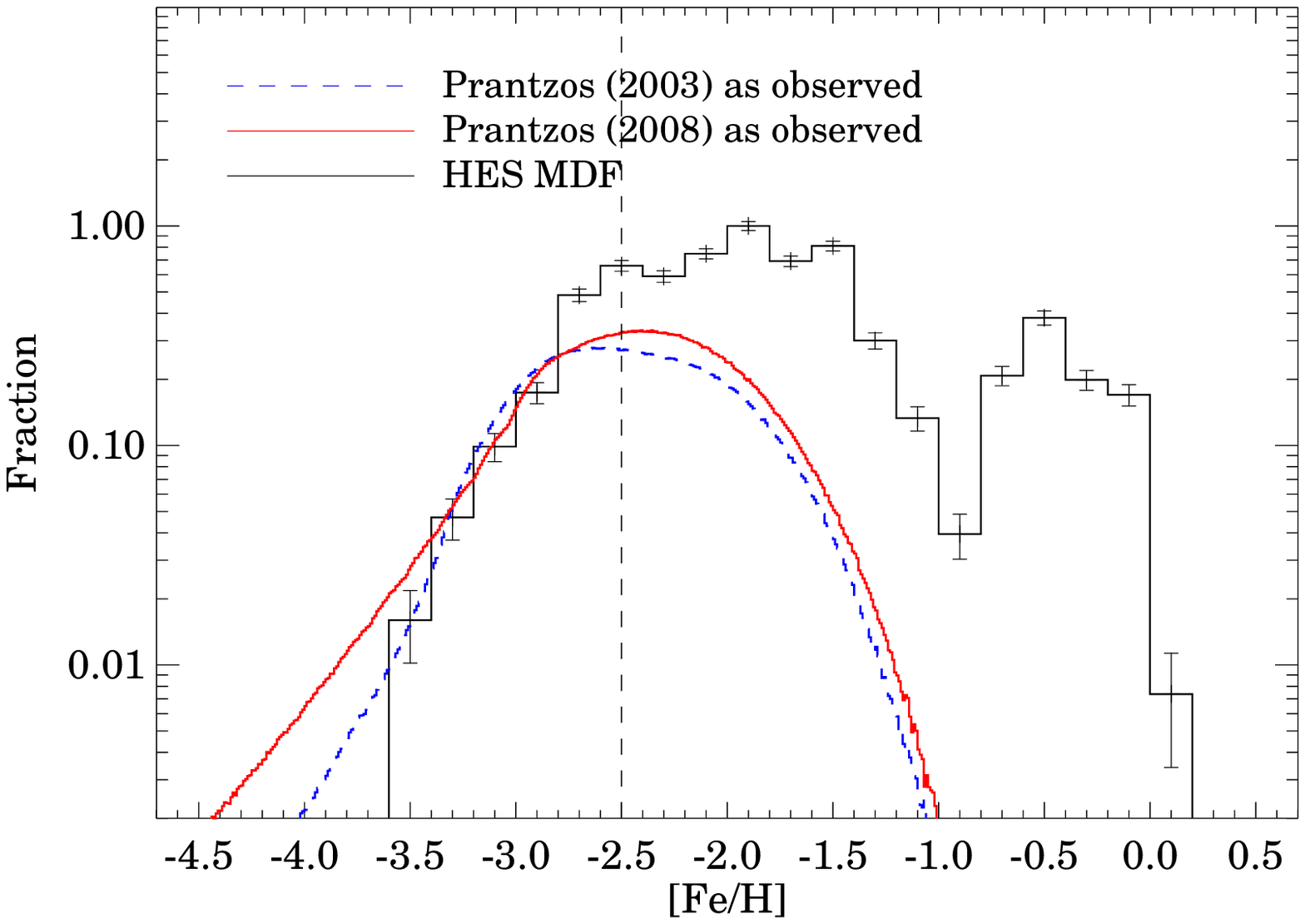}
\qquad
\includegraphics[width=0.46\textwidth]{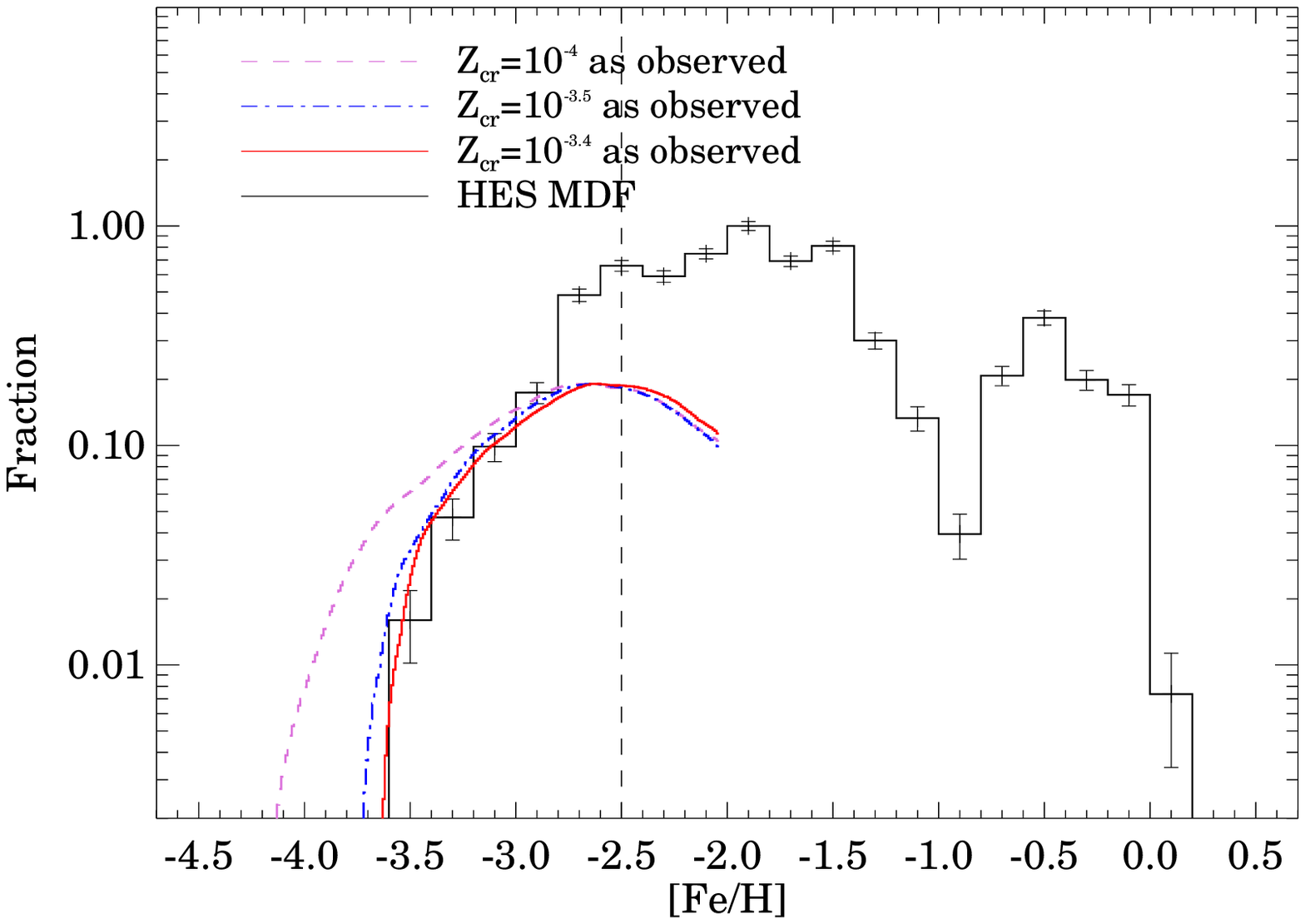}
\caption{Comparison of the MDF of main sequence and turn-off halo stars of the HES sample
to theoretical
predictions {\bf Left:}. Comparison to Prantzos (2003b) and Prantzos(2008a). {\bf Right:}
Comparison to Salvadori et al. (2007). Figure from Li et al. (2010).   }
\label{fig:4}
\end{figure}

%The lower mass limit $M_1$ is adopted here to be $M_1$=1-2 10$^6$ \ms, in
%agreement with the lower mass bound of observed dSphs in the Local group.
% Such galaxies have internal velocities $V>$10
%km/s. Dekel and Woo (2003) argue that the gas in haloes with
%$V<$10 km/s cannot cool to form stars at any early epoch and that dwarf
%ellipticals form in haloes with 10$<V<$30 km/s, the upper limit corresponding to 
%a stellar mass $M_*\sim$2 10$^8$ \ms. This is the 
%upper mass limit $M_2$ that we adopt here. 
The main properties of the sub-halo set constructed in this section
appear in Fig. 5 (left) as a function of the stellar sub-halo mass $M_*$.
The resulting total DMD is obtained as a sum over all sub-haloes:
\begin{equation}
\frac{d(n/n_1)}{d(logZ)} \ = \ \int_{M_1}^{M_2} \ \frac{d[n(M_*)/n_1(M_*)]}{d(logZ)} \ \frac{dN}{dM_*} \ M_* \ dM_*
\end{equation}
The result appears in Fig. 5 (right, with top panel in linear and middle panel in logarithmic scales, respectively). It can be seen that it fits the observed DMDs at least as well as the simple model \`a la Hartwick. 
In summary, under the assumptions made here, the bulk of the
DMD of the MW halo results naturally as the sum of the DMDs of the
component sub-haloes and can be understood analytically.
 It should be noted that all the ingredients of the analytical model are taken from observations of
local satellite galaxies, except for the adopted mass function of the sub-haloes (which results from analytical theory of structure formation plus a small modification to 
account for the role of outflows). 
%Obviously, by assuming different values 
%for the slope of the dark matter haloes  %and for the mass limits $M_1$ and $M_2$ 
%one may modify the peak of the resulting composite DMD, thus allowing for differences 
%between the halo DMDs of different galaxies.

Besides the shape of the bulk of the halo DMD, its low metallicity tail offers valuable
clues as to the early period of halo formation and metal enrichment. Recent analysis
of the HES data, for $\sim$1700 giant stars (Sch\"orck et al. 2009) and for $\sim$700 turn-off
stars (Li et al. 2010, Fig. 6) 
seem to suggest a sharp decline in star numbers below [Fe/H]$\sim$--3.5,
which could be interpreted as evidence for halo formation from
gas pre-enriched to that value (e.g. Salvadori et al. 2007). However,  the
situation may be more complex ("dual" halo structure, with unknown relative contributions from
an inner, metal-rich and an outer, metal-poor halo, Carollo et al. 2007) and small
number statistics at such low metallicities prevent any definitive conclusions yet.
Fortcoming studies (SEGUE-2, APOGEE, LAMOST) are expected to clarify the situation
in that metallicity range.

\section{Radial mixing in the Milky way disk}

In classical studies of GCE it is explicitly assumed that the system may be "open" as far as its
gas is concerned (allowing for e.g. infall, outflow or radial inflows) but it is "closed" regarding
its stars: once formed they remain in the system and their properties (especially those
of long-lived ones: metallicity distribution, age-metallicity relation) can help us to
reconstruct the history of the system. This paradigm started changing in recent years,
making the interpretation of stellar data more difficult (requiring combined studies
of chemistry and kinematics), but also more enriching, opening new perspectives.

The idea that stars in a galactic disk may diffuse to large distances along the radial 
direction (i.e to distances larger than allowed by their epicyclic motions) was proposed 
by Wielen et al. (1996). They suggested that some of the peculiar chemical properties 
of the Sun may be explained by the assumption that it was born in the inner Galaxy 
(i.e. in a high metallicity region, in view of the galactic metallicity 
gradient) and subsequently migrated outwards. They treated the hypothetical radial 
migration phenomenologically, acknowledging that the basic mechanism 
for the gravitational perturbations of stellar orbits is not understood.

Sellwood and Binney (2002, herefter SB02) convincingly argued that stars can 
migrate over large radial distances,
due to continuous resonant interactions with transient 
spiral density waves at co-rotation. Such a migration alters 
the specific angular momentum of individual stars, but affects very little the overall 
distribution of angular momentum and thus does not induce important radial heating of the disk.
Because high-metallicity stars from the inner (more metallic and older) and the outer 
(less metallic and younger) disc are brought in the solar neighborhood, SB02 showed 
with a simple toy model that considerable scatter may result in the local age-metallicity
relation, not unlike  the one observed by Edvardsson et al. (1993); see also Prantzos (2008b). 

Another obvious implication of the radial migration model of SB02 concerns the flattening 
of the stellar metallicity gradient in the galactic disk. That issue was quantitatively 
explored in Lepine et al. (2003), who  considered, however, the corotation at a fixed radius 
(contrary to SB02). As a result, the gravitational interaction bassically removes stars 
from the local disk, "kicking" them inwards and outwards. The abundance profile 
(assumed to be initially exponential) is little affected in the inner Galaxy, but some
 flattening is obtained in the 8-10 kpc region. The authors claim that such a 
 flattening is indeed observed (using data of planetary nebulae by Maciel and Quireza 1996) but modern surveys do not find it.

\begin{figure}
\begin{center}
\includegraphics[width=0.96\textwidth]{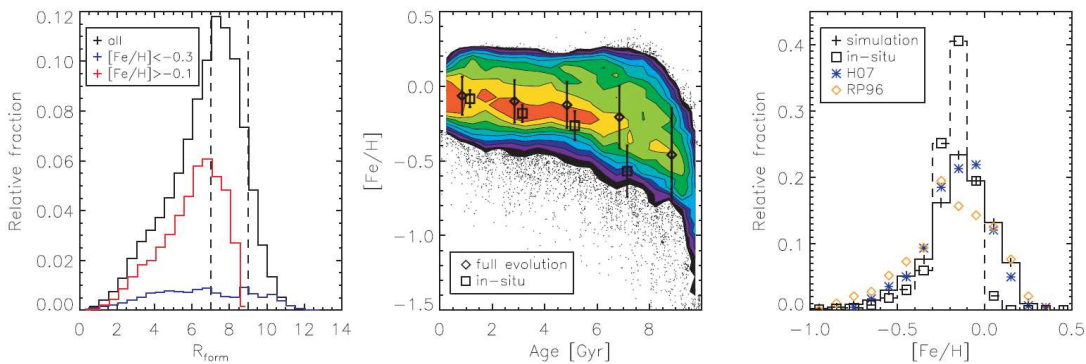}
\caption{Properties of stars in the solar neighborhood. {\bf Left}: Histogram of birth radii for stars that end up in the solar neighborhood on nearly circular orbits.
The black, red, and blue lines represent all, metal-rich, and metal-poor stars, respectively. {\bf Middle}: The age-metallicity relation: color contours represent relative
particle densities where point density is high. Diamonds and error bars indicate mean values and dispersion, respectively. Squares show the AMR if stars are
assumed to remain in situ. A small horizontal offset is applied to the two sets of symbols for clarity. {\bf Right}: Metallicity distribution function (MDF): the simulated
distribution is shown with the solid black histogram; diamonds and asterisks show data from Rocha-Pinto and Maciel (1996) and Holmberg et al. (2007), respectively.
The dashed histogram is the MDF if stars are assumed to remain in situ
(from Roskar et al. 2008).   }
\label{fig:4}
\end{center}
\end{figure}

On the basis of kinematics and abundance observations of a 
large sample of local stars  Haywood (2008) argues that most of the metal rich stars in the solar
neighborhood originate from the inner disk and most of the metal poor ones from the outer disk, 
and suggests that the local disk started its evolution with a considerably high metallicity
of [Fe/H]$\sim$-0.2. However,  such a large pre-enrichment of the thin disk is difficult
to accept, because no other local component of the Galaxy is massive enough 
to  enrich to such a high level the massive thin disk. Independently, however,
of his far-reaching conclusions, Haywood (2008) presents convincing arguments that the local 
stellar population shows evidence for substantial 
contamination with stars from other
Galactic regions. This idea has profound implications for galactic chemical evolution studies,
since it implies that observations of a stellar population in a given region cannot be
used to derive the history of that region: the history of adjacent (and even remote) regions
has to be considered as well.

Building on the ideas of SB02, Schoenrich and Binney (2009, SB09) presented a full scale
semi-analytic model for the chemical evolution of the Milky Way disk, including
several ingredients: gaseous infall, radial inflow of gas along the disk, churning of stars and 
cold (but not hot) gas and blurring of stars\footnote{In SB terminology, {\it churning} implies change of
guiding-centre radii while {\it blurring} means steady increase of the oscillation amplitude
around the guiding-centre, boths effects being due to interaction with spiral arm potential.}.
The model has a rather large number of parameters and assumptions 
and finds excellent agreement with 
each and every  observable in the solar neighborhood (including shape and scatter in 
age-metallicity relation, G-dwarf metallicity distribution, kinematics of thin and thick
disk etc.). In particular, the properties of the thick disk are "naturally" found in this model
as a result of secular evolution, with no need to invoke galactic mergers.

Numerical (N-body + SPH) simulations of Roskar et al. (2008) have already shown
that extensive radial mixing may occur in disk galaxies, due to the action
of spiral arms, and that it may help explaining observed properties 
of the solar neighborhood (Fig. 7). Recent simulations of Loebman et al. (2010) lend support to the idea
of thick disk resulting from secular evolution
(albeit with substantial differences on some observables with respect to SB09): the local thick
disk results from stars migrated from the inner disk,  retaining their (high) vertical velocity
dispersions but found in  the lower gravitational potential of the solar neighborhood.
Finally, Minchev and Famaey (2010) find that the galactic bar, in conjunction with the spiral arm 
potential, may play an efficient role in accelerating radial migration of stars.

Although it is rather early to say whether the global picture of the Milky Way evolution 
(involving an inside-out disk formation) will change drastically, it is clear that those works open new and promising perspectives in GCE studies. 

\medskip

\noindent{\bf Acknowledgements}: I am grateful to the organizers of NiCXI for their invitation and financial
support.

\end{document}